\documentclass[aps,prl,twocolumn,showpacs,floatfix]{revtex4}
\usepackage{epsfig}

\begin{document}
\title{Extended Self-Similarity and Hierarchical Structure in Turbulence}
\author{Emily S. C. Ching$^1$, Zhen-Su She$^{2,3}$,
Weidong Su$^{1,2}$ and Zhengping Zou$^2$}
\affiliation{$^1$Department of Physics, The Chinese University of Hong Kong,
Shatin, Hong Kong \\
$^2$ State Key Laboratory for Turbulence Research. Dept. of Mech.
and Engin. Sci., Peking University,
Beijing 100871, People's Republic of China \\
$^3$ Department of Mathematics, UCLA, Los Angeles, California 90095}
\date{\today}
\begin{abstract}
It is shown that the two remarkable properties of turbulence,
the Extended Self-Similarity (ESS) [R. Benzi {\it et al.}, Phy. Rev. 
E {\bf 48}, R29, (1993)] and
the She-Leveque Hierarchical Structure (SLHS) 
[Z.S. She and E. Leveque, Phy. Rev. Lett.
{\bf 72}, 336, (1994)] are
related to each other. In particular, we have shown that
a generalized hierarchical structure together with the most intense 
structures being shock-like give rise to ESS. Our analysis thus suggests
that the ESS measured in turbulent flows is an indication of the shock-like
intense structures. Results of analysis of velocity measurements in a pipe-flow
turbulence support our conjecture.
\end{abstract}
\maketitle

Fully developed turbulence is characterized by
power-law dependence of the moments of velocity fluctuations. 
It was suggested by Kolmogorov in 1941 (K41)\cite{K41} 
that there is a constant rate of energy transfer 
from large to small scales and that the statistical properties 
of the velocity difference 
across a separation $r$, $\delta v_r \!\equiv\! v(x+r)\!-\!v(x)$, depend
only on the mean energy transfer or equivalently the mean energy 
dissipation rate $\epsilon$ and the scale $r$ when $r$ is within an 
inertial range. Dimensional considerations then lead to the prediction 
that the velocity structure functions, which are moments of the magnitude
of the velocity difference, have simple power-law dependence on $r$ within
the inertial range:
\begin{equation}
S_p(r) \equiv \langle |\delta v_r|^p \rangle \sim \epsilon^{p/3} r^{p/3}
\label{K41}
\end{equation}

Experiments\cite{Frisch} have indicated that there is indeed power law
scaling in the inertial range but the scaling exponents are different from
$p/3$:
\begin{equation}
S_p(r) \sim r^{\zeta_p}
\label{inter}
\end{equation}
where $\zeta_p$ has a nonlinear dependence on $p$.
Such a deviation implies that the functional form of the 
probability density function (pdf) of $\delta v_r$ depends on $r$, 
that is, the velocity fluctuations have scale-dependent statistics.
Understanding this deviation from K41 is essential to our 
fundamental understanding of the small scale statistical 
properties of turbulence.

Recently, Benzi {\it et al.}\cite{ESS} have discovered a
remarkable new scaling property: $S_p(r)$ has a power-law
dependence on $S_3(r)$ over a range substantially longer than
the scaling range obtained by plotting $S_p(r)$ as a function of $r$. 
This behavior was named {\it Extended Self-Similarity} (ESS); its
discovery has enabled more accurate determination of the 
scaling exponents $\zeta_p$, particularly at moderately high Reynolds 
numbers assessible experimentally and numerically, 
It was later reported that ESS is invalid for
anisotropic turbulent flows such as atmospheric boundary layer and
channel flow\cite{SS93,BST96,ABS97}. This inspires the study of
a generalized ESS (GESS), a scaling behavior of
the normalized structure functions when plotted against each
other\cite{BBCST95,BBCST96}, which is still valid in these anisotropic
flows. The validity of ESS suggests that the different order 
structure functions have the same dependence on $r$ when 
$r$ is near the dissipative range\cite{BCBC95,SLP96,FG01}.
Very recently, Yakhot argued that some mean-field approximation of
the pressure contributions in the Navier-Stokes equation would
lead to ESS\cite{Yakhot}.
 
A number of phenomenological models have been proposed to 
explain the anomalous scaling exponents $\zeta_p$. 
Among them, a recent successful one 
is from She and Leveque\cite{SL94} with a hypothesis of 
a hierarchical structure (HS). When stated for the velocity structure
functions, the HS hypothesis reads
\begin{equation}
{S_{p+2}(r) \over S_{p+1}(r)} = A_{p+1} \left[{S_{p+1}(r) \over
S_{p}(r)}\right]^{\beta} [S^{(\infty)}(r)]^{1-\beta}
\label{SLvelocity}
\end{equation}
Here $S^{(\infty)}(r) \!\equiv\! \lim_{p \!\to\! \infty} S_{p+1}(r)/ S_{p}(r)$
and $0 \!<\! \beta \!<\! 1$ is a constant. 
This hypothesis (and the similar version for the local energy dissipation) 
was supported by experimental velocity measurements
taken in turbulent jets and wakes\cite{CBC95,CBBC95,CB97}.
Note that the boundedness of the velocity will ensure the existence of
$S^{(\infty)}(r)$. In fact, one can show that 
$S^{(\infty)}(r) \!\equiv\! \lim_{p \to \infty} S_p^{1/p}$, and is equal to
$|\delta v_r|^{\max}$,
the maximum magnitude of $\delta v_r$\cite{ChingSu}.

It was later reported that
the passive scalar structure functions\cite{CBC96} and
the local passive scalar dissipation\cite{LRCBC99},
the temperature structure functions\cite{Ching00}
and the local temperature dissipation\cite{ChingKwok00} 
in turbulent convection, and the velocity structure 
functions in a class of shell models\cite{FDB95,BBT96,LS97} 
were all found to possess similar hierarchical structures.
 
In this Letter, we shall show that the She-Leveque hierarchical 
structure (SLHS) leads to GESS. In other words, the SLHS is 
a special form of GESS. We then give a generalized form of SLHS,
which is equivalent to GESS, with the constant $\beta$ replaced 
by a function of $p$. Furthermore, it is shown that if 
$|\delta v_r|^{\max}$ is independent of $r$ for $r$ within the GESS
range (which will be the case when the most intense structures 
are shock-like), the generalized SLHS (and hence GESS) will give 
rise to ESS. We then conjecture that the observed ESS in turbulent
flows is an indication of the most intense structures
being shock-like. Consequently, we predict that in the anisotropic flows where
GESS but not ESS holds, $|\delta v_r|^{\max}$ has a dependence on $r$.
Finally, we present a systematic procedure of analysis
of experimental turbulent signals. Application of
this analysis to turbulent velocity fluctuations in a pipe flow
demonstrates that $|\delta v_r|^{\max}$ indeed depends on $r$ 
in the near-wall strong-shear regions where only GESS but not ESS is valid
but is consistent with being $r$-independent in the centerline of the pipe
where ESS is valid.
 
First, rewrite Eq.~(\ref{SLvelocity}) as
\begin{equation}
{S_{k+1}(r) \over S_{k}(r) S^{(\infty)}(r)} = A_{k} 
\left[{S_{k}(r) \over S_{k-1}(r) S^{(\infty)}(r)}\right]^{\beta}
\label{rewriteSL}
\end{equation}
for integer $k$, which implies
\begin{equation}
{S_{k}(r) \over S_{k-1}(r) S^{(\infty)}(r)} = 
\Pi_{j=0}^{k-1} A_j^{\beta^{k-1-j}}
\left[{S_1 (r) \over S^{(\infty)}(r)} \right]^{\beta^{k-1}}
\label{step1}
\end{equation}
ultiply Eq.~(\ref{step1}) for $k = 1, 2, \ldots, n$ gives
\begin{equation}
S_n(r) = B_n
\left[ {S_1(r) \over S^{(\infty)}(r)} \right]^{1-\beta^n \over 1-\beta}
[S^{(\infty)}(r)]^n
\label{step2}
\end{equation}
where $B_n \!\equiv\! \Pi_{k=1}^n \Pi_{j=0}^{k-1} A_j^{\beta^{k-1-j}}$,
which further gives
\begin{equation}
{S_n(r) \over S_3(r)^{n/3}} = {B_n \over B_3^{n/3}}
\left[ {S_1(r) \over S^{(\infty)}(r)} \right]^
{3(1-\beta^n)-n(1-\beta^3) \over 3(1-\beta)}
\label{step3}
\end{equation}
Equation~(\ref{step3}) then implies the GESS property:
\begin{equation}
T_n(r) \sim T_m(r)^{\rho(n,m)}
\label{GESSt}
\end{equation}
which is a scaling behavior for the normalized structure functions, 
$T_n(r) \!\equiv\! S_n(r)/S_3(r)^{n/3}$, with the normalized
exponents $\rho(n,m)$ even when $S_n(r)$ does 
not have a scaling behavior in $r$. For SLHS, 
\begin{equation}
\rho(n,m)= {3(1-\beta^n)-n(1-\beta^3) \over 3(1-\beta^m)-m(1-\beta^3) }
\label{rhoSL}
\end{equation}
We have thus proved that SLHS implies GESS. 
 
Earlier, Benzi et al.\cite{BBCST96} showed that
GESS holds if the structure functions are of the form
\begin{equation}
S_p(r) \sim g_1(r)^p g_2(r)^{H(p)}
\label{g1g2}
\end{equation}
for any functions $g_1(r)$, $g_2(r)$, and $H(p)$. On the other hand,
if Eq.~(\ref{GESSt}) with any $\rho(n,m)$ is valid, 
we can always write the structure functions in the form of 
Eq.~(\ref{g1g2}), say, with $g_1(r) = S_3(r)^{1/3}$,
$g_2(r)=S_{q^*}(r)/S_3(r)^{{q^*}/3}$, and $H(p) = \rho(p,q^*)$ 
for any chosen value of $q^*$. Thus, Eq.~(\ref{g1g2}) is 
equivalent to GESS.

Our demonstration here gives the functions $g_1(r)$, $g_2(r)$ 
a meaning. Although the choice of $g_1(r)$ and $g_2(r)$ is
not unique, the following form can be obtained from Eq.~(\ref{step2}):
$g_1(r)=S^{(\infty)}(r)$ and $g_2(r)=S_3(r)/[S^{(\infty)}(r)]^3$.
Here, $g_1(r)$ describes the $r$-dependence of very strong fluctuations
$S^{(\infty)}(r)$ and $g_2(r)$ describes the normalized $r$-dependence of 
(typical) weak fluctuations [e.g. $S_3(r)$] by $[S^{(\infty)}(r)]^3$. 
Express $g_1$ and $g_2$ this way, we have
\begin{equation}
S_p(r) \sim \left[S^{(\infty)}(r)\right]^p 
\left\{{S_3(r) \over [S^{(\infty)}(r)]^3}\right\}^{f(p)}
\label{modg1g2}
\end{equation}
where $f(p)$ is a function to be discussed below. Note that
when $g_2(r)$ is constant, or the weak fluctuations have the
same $r$-dependence as the strong fluctuations, we have the K41
scaling. The function $f(p)$ has to satisfy various conditions.
By definition, $f(0)=0$, $f(3)=1$, and 
$\lim_{p\to \infty} f(p+1)-f(p)=\lim_{p\to \infty}f(p)/p = 0$. 
Furthermore, the boundedness 
of the velocity restricts that $df(p)/dp \ge 0$. 
 
Note that Eq.~(\ref{modg1g2}) is a general expression of GESS.
Rewriting Eq.~(\ref{modg1g2}) in the form of Eq.~(\ref{rewriteSL}),
we have 
\begin{equation}
{S_{p+1}(r) \over S_{p}(r) S^{(\infty)}(r)} = C_{p}
\left[{S_{p}(r) \over S_{p-1}(r) S^{(\infty)}(r)}\right]^{g(p)}
\label{generalHS}
\end{equation}
where $g(p) \equiv [f(p+1)-f(p)]/[f(p)-f(p-1)]$.
It is clear that SLHS corresponds to the
particular case of $g(p)= \beta$ or equivalently 
\begin{equation}
f(p)={1-{\beta}^p \over 1-\beta^3}
\label{fpSL}
\end{equation}
We can list several additional possible classes of $f(p)$, which
include $f(p)\!=\! [(p\sigma+1)^{\alpha} \!-\! 1]/
[(3\sigma+1)^{\alpha} \!-\! 1]$ and
$f(p) \!=\! \ln(p \sigma+ 1)/\ln(3 \sigma + 1)$
with $\sigma \!>\! 0$ and $0\!<\! \alpha\!<\! 1$. 
The latter is the limit of the former when $\alpha \to 0$. 
These two cases correspond to the two results derived by
Novikov, Eqs. (18) and (19) in Ref.\cite{N94}, using the 
theory of infinitely divisible distributions. 
 
To study when GESS would further give rise to ESS,
we rewrite Eq.~(\ref{GESSt}) in the following form:
\begin{equation}
{S_p(r)^{1/p} \over S_3(r)^{1/3}} \sim 
\left[{S_q(r)^{1/q} \over S_3(r)^{1/3}}\right]^{{q \over p} \rho(p,q)}
\label{rewriteGESS}
\end{equation}
with 
\begin{equation}
\rho(p,q) = {f(p) - p/3 \over f(q)- q/3}
\label{rho}
\end{equation}
Thus we see that if there exists an $p^* \ne 0$ such that 
$S_{p^*}(r)^{1/p^*}$
is independent of $r$, then we would have a scaling behavior of $S_p$ vs
$S_3$ (and thus $S_q$ for $q \ne p$). 
If $p^*$ is finite, $\zeta_{p^*} = 0$ which
implies $\zeta_m = 0$ for $0 \le m \le p^*$ because $\zeta_n$ has
to be a non-decreasing function of $n$ in order for the velocity field
to be bounded\cite{Frisch}. It further gives $\zeta_p = 0$ for all
values of $p$ if $\zeta_p$ is an analytic function of $p$. 
This could be avoided only if $p^*$ approaches $\infty$, 
which means that $S^{(\infty)}(r)$ is independent of $r$.
Hence, GESS together with the condition that 
$S^{(\infty)}(r)$ or equivalently $|\delta v_r|^{\max}$ 
being independent of $r$ would give rise to ESS:
\begin{equation}
S_p(r) \sim S_3(r)^{\eta(p,3)}
\label{ESSt}
\end{equation}
with $\eta(p,3)\!=\! f(p)$
even when $S_p(r)$ does not have a power-law dependence on $r$. 
We note that the presence of shocks with maximum velocity, $v_{\max}$,
in opposite directions across the shock discontinuity would give
$|\delta v_r|^{\max} = 2 v_{\max}$, which is thus independent of $r$.

Using Eq.~(\ref{modg1g2}), an independence of $S^{(\infty)}$ on $r$ implies 
\begin{equation}
\zeta_p = \zeta_3 f(p)
\label{zetap}
\end{equation}
For SLHS, $\lim_{p \to \infty} f(p) \!=\! 1/(1-\beta^3)$ hence 
Eq.~(\ref{zetap})
further implies a saturation of $\zeta_p$ as $p \to \infty$.
A connection of the saturation of the exponents with the existence of
shocks was suggested earlier by Chen and Cao\cite{CC95}.

Both ESS and SLHS have been observed in a variety of fluctuating 
flow fields. The above analysis points to a possibility that
the two combined may be indicative of a property of the flow field:
$S^{(\infty)}(r)$ is independent of $r$. One way to  
check the plausibility of these ideas is to examine experimental 
velocity measurements that demonstrate GESS, and to investigate 
whether $|\delta v_r|^{\max}$ is indeed independent of $r$ when
ESS is valid and otherwise dependent on $r$ when ESS is not valid.
Note, however, that the detectable $|\delta v_r|^{\max}$ 
in any finite measurement would almost surely underestimate 
the true value. 

If the GESS is of the special form of SLHS, a systematic procedure
can be developed to perform an indirect estimate of the
$r$-dependence of $S^{(\infty)}(r)$ as follows.
First, one verifies if the SLHS is valid by performing
a $\beta$-test. It consists of computing the normalized structure
functions $T_p(r)$ and obtaining the normalized exponents
$\rho(p,q)$ by measuring the slopes of $\ln(T_p)$ vs 
$\ln(T_q)$. Let $\Delta\rho(p,q)=\rho(p+1,q)\!-\!\rho(p,q)$.
It is easy to derive the following equation when SLHS is valid:
\begin{equation}
\label{beta-test}
\Delta\rho(p+1,q) =\beta \Delta\rho(p,q)
\!+\!{(1-\beta)(1-\beta^3)\over q(1-\beta^3)\!-\!3(1-\beta^q)}
\end{equation}
If one finds parallel straight lines when
ploting $\Delta\rho(p+1,q)$ vs $\Delta\rho(p,q)$
for a set of values of $q$, 
we say that the data passes the $\beta-$test and the turbulent 
flow field possesses the SLHS property. The slope and intercept 
provides a double estimate of the constant $\beta$. 
With the estimated $\beta$,
one can then construct $f(p)$ using Eq.~(\ref{fpSL}).

For an indirect estimate of $S^{(\infty)}(r)$, we introduce 
\begin{equation}
\label{Fp}
F_p(r,r_0)\!\equiv\! {\ln[S_p(r)/S_p(r_0)] \!-\! f(p)
\ln[S_3(r)/S_3(r_0)] \over p-3f(p)}
\end{equation}
From Eq.~(\ref{modg1g2}), 
$F_p(r,r_0)$ should be independent of $p$ and
equal to $\log [S^{(\infty)}(r)/S^{(\infty)}(r_0)]$ for 
$r$ within the GESS range. 
Thus, one computes $F_p(r,r_0)$ for a fixed value of $r_0$,
within the GESS range, and plots it as a function of 
$r$ for a set of values of $p$ to estimate indirectly the 
$r$-dependence of $\ln [S^{(\infty)}(r)/S^{(\infty)}(r_0)]$.
In particular, if $S^{(\infty)}$ is independent of $r$, then
$F_p(r,r_0)=0$ for $r$ within the GESS range.
 
\begin{figure}
\centering
\includegraphics[width=.40\textwidth]{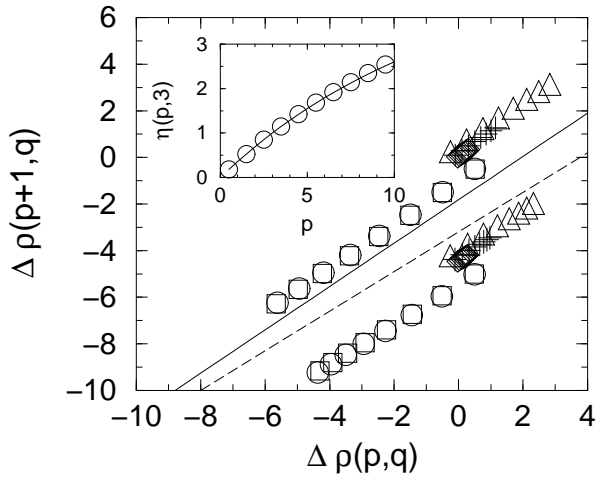}
\caption{
$\Delta \rho(p+1,q)$ vs $\Delta \rho(p,q)$ 
for $q\!=\!1$ (squares), $q\!=\!2$ (circles), $q\!=\!4$~(triangles), 
$q\!=\!5$~(plusses), and $q\!=\!8$ (diamonds).
Above the solid line (slope $\!=\! 0.93$) are the
centerline data and below the dashed 
line (slope $\!=\! 0.85$) are the near-wall data (shifted for clarity). 
In the inset, we compare $\eta(p,3)$~(circles) with
$f(p)$~(solid line) for the centerline data.}
\label{rhoplot}
\end{figure}

We have applied the above procedure to analyze hot-wire measurements of 
longitudinal velocity fluctuations in a pipe flow\cite{note}. 
The pipe is 22.5~m long with an inner diameter of 10.5~cm, and
the Reynolds number is $1.35 \times 10^5$.
The velocity measurements studied were taken at 
a cross section at 18.2~m away from the entrance both at the
centerline of the pipe and at a distance of 0.1~mm from the pipe-wall.
The number of measurements taken at each point is $5.76 \times 10^7$ 
and the maximum order of moments computed is $p=10$.
We have checked that GESS is valid for both sets of measurements but
ESS is valid only at the centerline.
The range of GESS (and also ESS when valid) is $r \approx 10 - 500$,
in units of the sampling time $1/48$~ms. All $r$'s quoted below will be
in the same units.
We have obtained $\rho(p,q)$ for the two locations and $\eta(p,3)$ for
the centerline location. In Fig.~\ref{rhoplot}, we plot
$\Delta \rho(p+1,q)$ vs $\Delta \rho(p,q)$
for some values of $q$ for both locations. 
The data can be fitted by parallel straight lines
showing that the hierarchical structure is indeed of the SL
form in both locations. 
We estimate the value of $\beta$ simultaneously from the slope
and the intercept and get $0.93 \pm 0.01$ 
and $0.85 \pm 0.01$ respectively for the centerline and the near-wall location. 
In the inset, we see good agreement of $\eta(p,3)$ with $f(p)$
for the centerline measurements. This implies the $r$-independence 
of $S^{(\infty)}$, in accord with the validity of ESS in these
measurements.

\begin{figure}
\centering
\includegraphics[width=.40\textwidth]{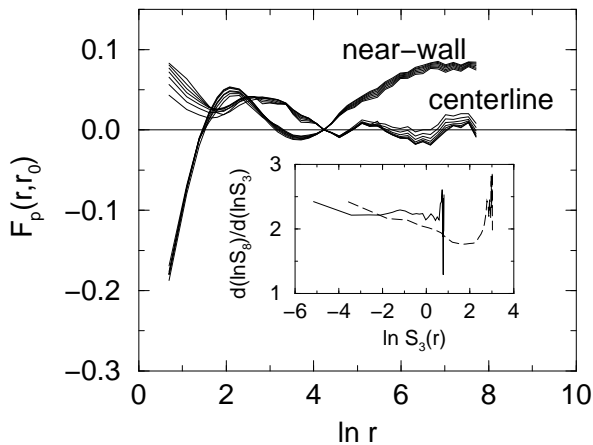}
\caption{$F_p(r,r_0)$ for $p \ge 4$ for both the centerline and near-wall
measurements. In the inset, we show the local slope of 
$\ln S_8(r)$ vs $\ln S_3(r)$ as a function of $\ln S_3(r)$, illustrating
that ESS is valid for the centerline (solid line)
but not the near-wall (dashed line) measurements.}
\label{Fplots}
\end{figure}

We have next computed $F_p(r,r_0)$ with $r_0\!=\!69$ for both locations
and found that the data indeed collapse when $p$ is large. 
In Fig.~\ref{Fplots}, we plot $F_p(r,r_0)$ as 
a function of $r$.
We see that $F_p(r,r_0)$ for the centerline measurements is almost
zero for $r \!\ge\! 30$ while that for the near-wall measurements
shows a clear $r$-dependence throughout. Since ESS is valid for the
centerline but not the near-wall measurements (see the inset), these
results support our conjecture that the validity of ESS in turbulent
flows is an indication of $S^{(\infty)}$ being independent of $r$.

In summary, our analysis proposes a link between a measurable
statistical property of turbulence, the ESS, and a property of
the most intense structure, the $r$-independence of 
$|\delta v_r|^{\max}$ or $S^{(\infty)}(r)$. We have developed a
systematic procedure to test this conjecture
when the SL hierarchical structure holds. 
The analyis of experimental pipe flow data support our conjecture:
the near-wall strong-shear turbulence contains more complex structures while
the centerline fully developed turbulence has shock-like 
structure for $S^{(\infty)}(r)$.
A consequence of the $r$-independence of $S^{(\infty)}(r)$ is
the saturation of the scaling exponents of the velocity 
structure functions at very high orders. 
Further experimental and numerical tests are highly desirable.  

\begin{acknowledgments}
Z.Z. thanks Prof. M.D. Zhou and Y. Zhu for their help in 
conducting the pipe-flow velocity measurements. The work at the 
Chinese University of Hong Kong was supported by the 
Research Grants Council of the Hong Kong SAR, China 
(CUHK 4119/98P and CUHK 4286/00P) 
while that at Peking University was partially supported by a grant
from National Natural Science Foundation (NNSF) of China and from the
Ministry of Education of China.
\end{acknowledgments}

\end{document}